\documentclass[aps,prl,notitlepage,reprint,twocolumn,showpacs,superscriptaddress]{revtex4-2}

\usepackage{epsfig}
\usepackage{natbib}
\usepackage{ulem}
\usepackage{xspace}
\usepackage{xcolor}
\usepackage{graphicx}
\usepackage{amssymb}
\usepackage{cancel}
\usepackage{ulem}
\usepackage{lineno}
\usepackage{amsmath}
\usepackage{stackrel}
\usepackage{appendix}
\usepackage{url}
 \usepackage{enumitem}
 \usepackage{hyperref}
\usepackage{xcolor}
\usepackage{array}
\usepackage{booktabs}
\usepackage{multirow}

\hypersetup{
	colorlinks=true,
	linkcolor=blue,
	citecolor=blue,
	urlcolor=black
}

\begin{document}
\title{First passage times in compact domains exhibit bi-scaling}
\author{Talia Baravi}
\affiliation{Department of Physics, Institute of Nanotechnology and Advanced Materials, Bar-Ilan University, Ramat Gan 52900, Israel}
  \author{David A. Kessler}
           \affiliation{Department of Physics, Bar-Ilan University, Ramat Gan 52900, Israel}
\author{Eli Barkai}
\affiliation{Department of Physics, Institute of Nanotechnology and Advanced Materials, Bar-Ilan University, Ramat Gan 52900, Israel}

\begin{abstract}
The study of first passage times for diffusing particles reaching target states is foundational in various practical applications, including diffusion-controlled reactions. In large systems, first passage times statistics exhibit a bi-scaling behavior, challenging the use of a single time scale. In this work, we present a bi-scaling theory for the probability density function of first passage times in confined compact processes, applicable to both Euclidean and fractal domains and for diverse geometries. Our theory employs two distinct scaling functions: one for short times, capturing initial dynamics in unbounded systems, and the other for long times is sensitive to finite size effects. The combined framework is argued to provide a complete expression for first passage time statistics across all time scales. As our detailed calculations show, the theory describes various scenarios with and without external force fields, for active and thermal settings, and in the presence of resetting when a non-equilibrium steady state emerges.
\end{abstract}
\maketitle

\textit{Introduction}--
Diffusion-controlled reactions under confinement regulate many physical, chemical, and biological processes. In this context a key observable is the time it takes a particle diffusing in a medium to reach a target position \cite{redner2001guide,bray2013persistence,meyer2011universality,mirny2008cell,benichou2008optimizing,grigoriev2002kinetics,basu2018active,boyer2004modeling,jolakoski2023first,lanoiselee2018diffusion,shlesinger2006search,burioni2005random,guerin2016mean,mattos2014trajectory,grebenkov2023boundary,scher2023escape,dybiec2006levy,valenti2007hitting,padash2022asymmetric,Dahlenburg_2023,dubkov2023enhancement}. Two questions of broad relevance are (i) What is the dependence of the distribution of this first passage time (FPT) on the initial target-source separation? (ii) How does this distribution depend on the size of the confinement region? While foundational theories initially promoted the use of exponential statistics, later research showed that the distributions are non-exponential \cite{godec2016first,godec2016universal}, involving multiple time scales \cite{benichou2014first}. Further, the mean time is in many cases not a good diagnostic of the process, as it does not capture the typical events \cite{mattos2012first}. Finally, the distinction between compact (i.e., recurrent in the absence of confinement) and non-compact search is vital \cite{condamin2007first}. 
\par The observed limitations of existing mono-scaling theories, which propose a single time scale to characterize diffusion-controlled reactions, led us to a new study based on bi-scaling \cite{castiglione1999strong,rebenshtok2014infinite,gal2010experimental,afek2023colloquium}. Characterizing the statistical laws of the FPT for large system sizes requires two scaling functions. Focusing on compact exploration, this provides a solution to a long-standing problem: the evaluation of the FPT distribution and the corresponding moments. We quantify the statistics of the FPTs by the dependence of the moments $\langle t^q\rangle$ on the confinement length $L$, 
\begin{equation}\label{eq1}
    \langle t^q \rangle\propto L^{q\nu(q)}
\end{equation} 
 with $q\geq 0$. The mono-scaling hypothesis posits that $\nu(q)$ is constant, suggesting a unique scaling function across different system sizes. This would imply that the probability density function (PDF) of FPTs adheres to a simple scaling form $\eta(t) \sim h(t/\tau_L)/\tau_L$, where $\tau_L$ is a characteristic time scale set by $L$. Such a framework, if true, would greatly simplify data analysis, as it involves only a single scaling function.  However, empirical observations and theoretical analyses reveal that the reality is more complex, and the mono-scaling hypothesis does not universally apply. For instance, a study by Meyer et al. \cite{meyer2011universality} introduced such a mono-scaling solution for diffusion on fractals, offering an asymptotic expansion of the PDF. Nevertheless, this approach fails to fully capture the statistical nuances of the system; the resulting solution is not normalizable and thus represents only a partial view of the overall statistical behavior. The limitations of mono-scaling theories become evident when considering the varied behavior of FPT distributions demonstrating the need for a more adaptable and general framework.

\par Are there any inherent universal characteristics of the function $\nu(q)$? It is well known that the dynamics of diffusion processes can be characterized by several key exponents: the fractal dimension $d_f$, which quantifies the number of sites $N\propto r^{d_f}$ within a sphere of radius $r$; the walk dimension $d_w$, defined in the absence of absorption via the mean square displacement (MSD) as $\langle r^2 \rangle\propto t^{2/d_w}$ \cite{ben2000diffusion,yuste1995first}; and the persistence exponent \(\theta\), which describes the decay of the survival probability in a semi-infinite system $S(t)\propto t^{-\theta}$ \cite{majumdar1999persistence}. Specific values of these exponents are known for a large number of systems \cite{bray2013persistence} and will be provided below when we treat examples. We claim that, for compact diffusion in confined systems, the relation
\begin{equation}\label{eq:nuq}
    q\nu(q)=\begin{cases}
        0,&q<\theta \\ d_w(q-\theta),& q>\theta \, ,
    \end{cases}
\end{equation}
 provides a general connection between the scaling spectrum $q\nu(q)$, the persistence exponent $\theta$ and the walk dimension $d_w$. The small-fractional moments $q<\theta$ are insensitive to the system size, as can be seen from Eqs.\,(\ref{eq1},\ref{eq:nuq}). These moments are controlled by those trajectories where the particle reaches the target before hitting the boundary. A sharp transition is found when $q=\theta$, since for $q>\theta$, the boundary-hitting trajectories dominate the average. The emerging physical picture differs from that presented in the major review in the field, see Eq.\,(163) in \cite{benichou2014first}. Importantly, $\nu(q)$ is bi-linear; hence, we may construct a bi-scaling theory, as shown below. 
\begin{figure}[t]
    \centering
    \includegraphics[width=0.45\textwidth]{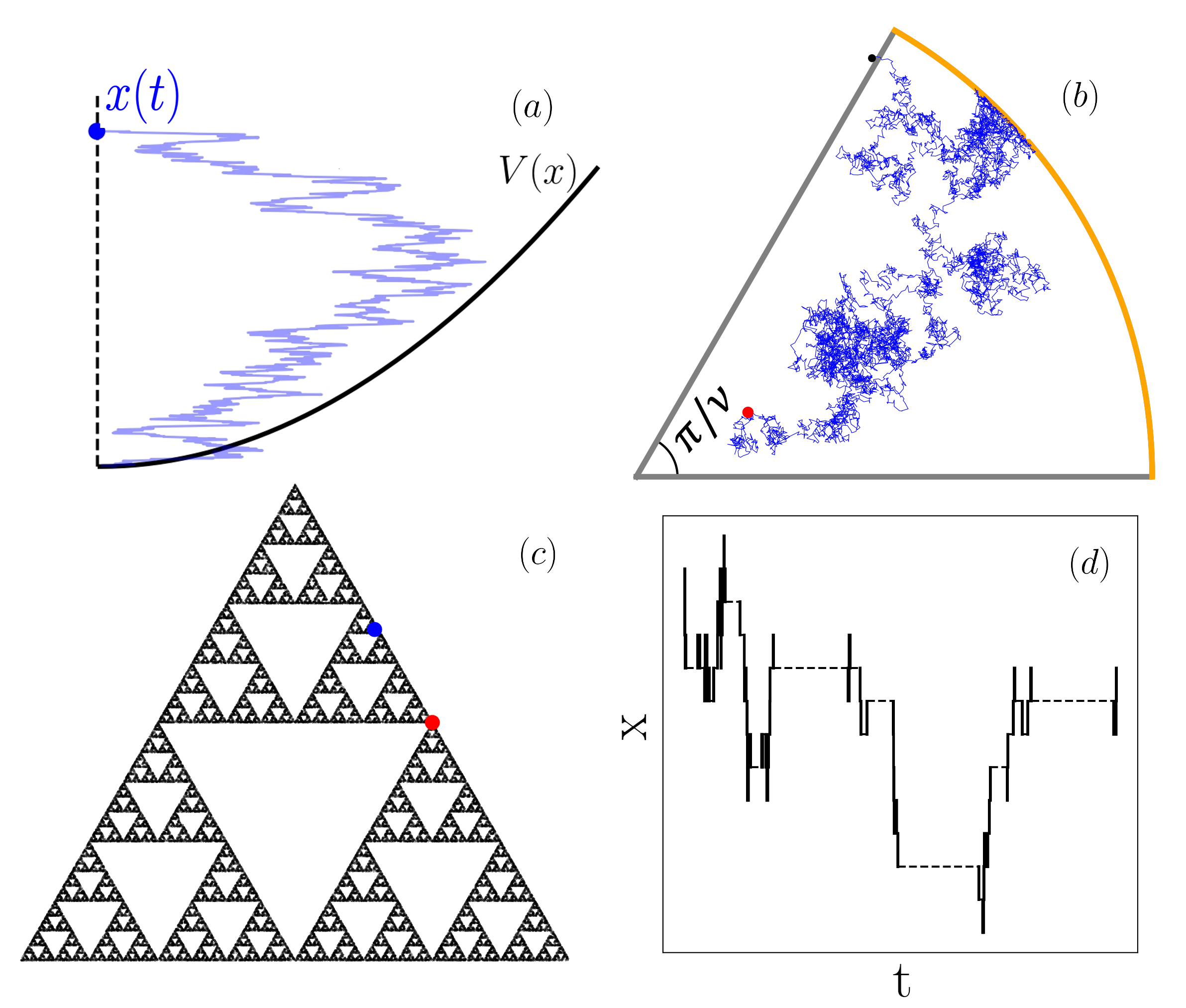}

    \caption{Models of FPT in confinement: (a) Brownian particle in Harmonic potential with an absorbing boundary at the bottom of the potential. (b) Trajectory (blue) of a random walker in a 2D wedge. The edges are absorbing. (c) Sierpinski gasket, where the blue dot is the initial position and the red dot is the absorbing boundary. (d) (x,t) diagram of one-dimensional CTRW process with long tailed waiting time PDF as found for diffusion in strongly disordered systems with trapping \cite{metzler1994fractional,kenkre1973generalized,metzler2000random}.  } 
    \label{fig:models}
\end{figure}
\par  Moreover, we claim that two scaling functions, $f_{\infty}(\cdot)$ and $\mathcal{I}(\cdot)$, are fundamental to understanding the problem. The first, $f_{\infty}(\cdot)$, varies with the initial state but is independent of system size, while the second function, $\mathcal{I}(\cdot)$, is unaffected by the initial state but depends on the system size. In the large \( L \) limit, where \(\tau_L\) diverges, the first-passage time distribution \(\eta(t)\) reveals a remarkable bi-scaling behavior: \(\eta(t) \sim f_{\infty}(t)\) for fixed \(t\) and \(\eta(t) \sim I(t/\tau_L)\) when \(t/\tau_L\) is fixed. Accurately solving first-passage time problems therefore requires evaluating both functions, an essential approach validated across diverse models in the accompanying paper \cite{Future_paper}. Critically, these solutions match at intermediate time scales, providing a consistent description across all scales.

\par More precisely, for compact diffusion processes in finite domains, in systems with or without an external force field, for diffusion on fractals and Euclidean spaces, for Markovian and semi-Markovian continuous time random walk (CTRW) presented schematically in Fig.\,(\ref{fig:models}), for systems with resetting or active models, the bi-scaling ansatz yields a general form for the PDF
\begin{equation}\label{eq:uniform2}
     \eta(t)\sim \left(\mathcal{N}/\tau_0\right)(t/\tau_L)^{1+\theta}f_{\infty}\left(t/\tau_0\right)\mathcal{I}\left(t/\tau_L\right) \, .
\end{equation}
In this context, $\tau_L\propto L^{d_w}$ represents the time scale associated with the size of the system, while $\tau_0\propto r_0^{d_w}$ is the time scale related to the initial distance from the target. This expression specifies how general claim that the FPT statistics for compact diffusion are described by two distinct functions $\mathcal{I}(\cdot)$ and $f_{\infty}(\cdot)$. These functions, which describe the short- and long-time behaviors of the FPT statistics, respectively, each exhibit a \( t^{-1-\theta} \) behavior for intermediate times. This feature holds for compact search, which enables the construction of the global solution presented in Eq.\,(\ref{eq:uniform2}). While rigorously proving universality is challenging, the consistent agreement of Eq.\,(\ref{eq:uniform2}) with results from nine diverse cases strongly supports its broad validity. The function $f_{\infty}(t)|_{t\gg 1}\propto t^{-1-\theta}$ is well studied, as it is the PDF of the FPT for infinite systems in the absence of external fields. The function $\mathcal{I}(\cdot)$ describes the tail characterizing the rare events. 
\par
\begin{figure}[t]
    \centering
  \includegraphics[width=0.45\textwidth]{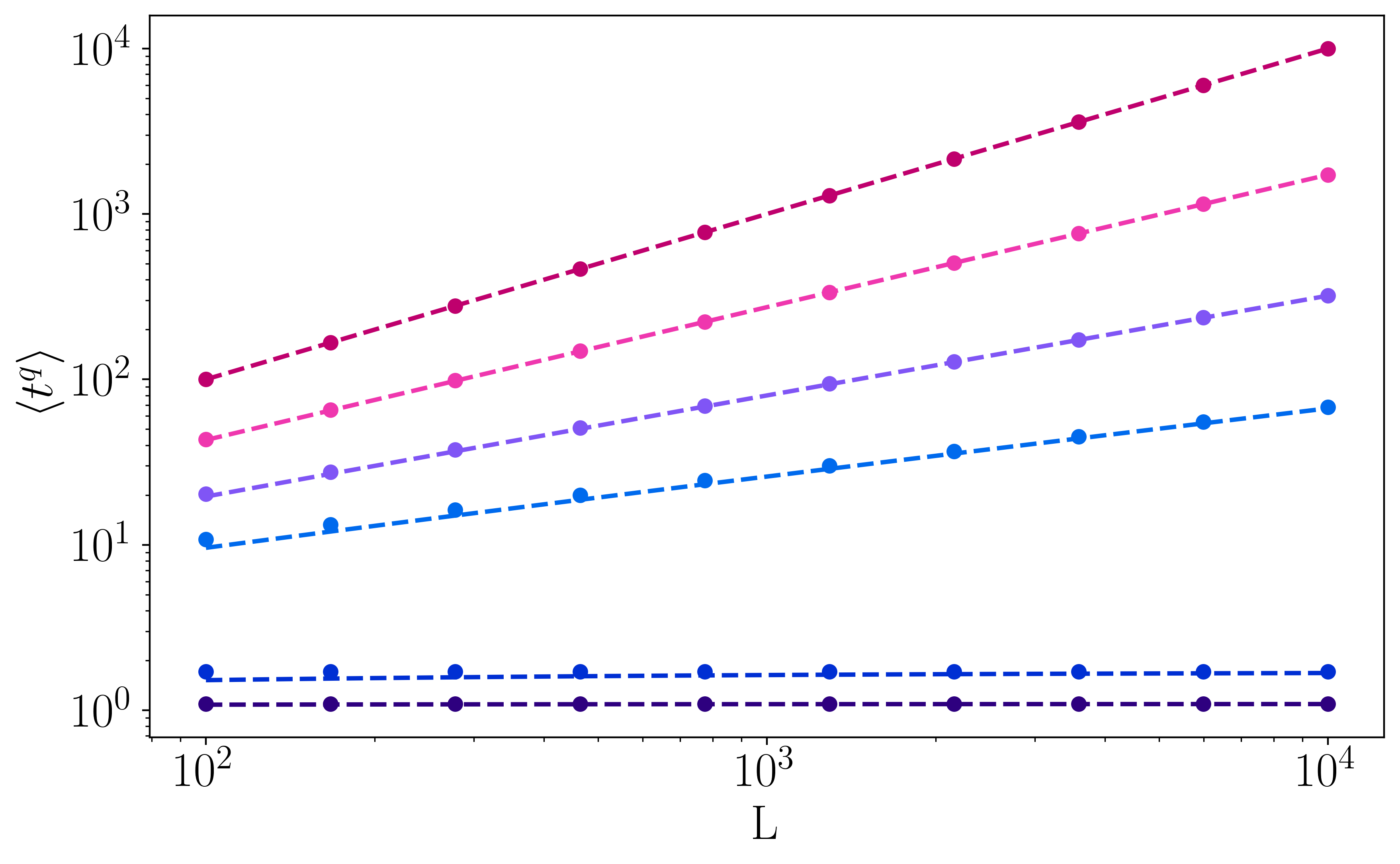}
    \caption{The FPT moments, $\langle t^q \rangle$, as a function of \( L = 2T / \kappa_1 \), where \( T \) denotes temperature, for \( q = 0.1, 0.3, 0.7, 0.8, 0.9, 1 \), corresponding to a particle confined in a V-shaped potential, \( V(x) = \kappa_1 |x| \). The dashed lines are the theory in Eq.\,(\ref{eq:nuq}) and the symbols are calculated by exact enumeration (see SM). From here we estimate the spectrum of the exponent defined in Eq. (\ref{eq:nuq}) with $\theta = 1/2$. The low order moments, $q<1/2$, show saturation effect, while for $q>1/2$, the moments increase with the confinement length $L$, as described by the scaling relationship in Eq. (\ref{generalscal1}).  } 
    \label{fig:sierpmom}
\end{figure}

 Our main focus lies on the function $\mathcal{I}(\cdot)$ and the use of Eq.\,(\ref{eq:uniform2}) in physical modeling. Mathematically, $\mathcal{I}(\cdot)$ is related to the PDF $\eta(t)$, an association that emerges through the long-time limit extrapolation from Eq.\,(\ref{eq:uniform2}) as
 \begin{equation}\label{definfhar}
\mathcal{I}(\tau)\equiv \lim_{t,\tau_L\rightarrow \infty}\tau_0^{-\theta}\tau_L^{1+\theta}\eta(t)\, ,
\end{equation}
with the ratio $\tau \equiv t/\tau_{L}$ maintained constant and $\eta(t)$ being the exact FPT PDF. Notably, since \(\eta(t)\) is normalized, it follows from the limit in Eq.\,(\ref{definfhar}) that \(\mathcal{I}(\tau)\) is non-normalizable. Therefore, this scaling solution does not represent a probability density.
As a result of Eqs.\,(\ref{eq1}-\ref{eq:uniform2}), the $q$-moment of the FPT exhibits two different asymptotic behaviors, given by 
\begin{equation}\label{generalscal1}
    \langle t^q\rangle\xrightarrow{}\begin{cases}
                     \int_0^{\infty}t^q f_{\infty}(t)dt , &  q<\theta\\
                     \tau_L^{q-\theta} \tau_0^{\theta} \int_0^\infty \tau^q \mathcal{I}(\tau)d\tau  &  q>\theta\\
                    \end{cases} \, .
\end{equation}
For $q<\theta$, the moments converge to the moments calculated from the PDF for an infinite system $f_{\infty}(t)$, and hence are $\tau_L$ independent. On the other hand, for $q>\theta$, the moments are given by the non-normalizable function $\mathcal{I}(\tau)$, and scale with $\tau_L^{q-\theta}$, which captures the long time limit of the FPTs. Eq.\,(\ref{generalscal1}) leads to the expression  
\begin{equation}\label{generalscalfrac}  
    \langle t^q\rangle \propto |q-\theta|^{-1}r_0^{qd_w-q\nu(q)}L^{q\nu(q)},  
\end{equation}  
where \(r_0\) represents the initial distance between the source and the target. Eq.\,(\ref{generalscalfrac}) indicates that the transition is characterized by a divergence of $\langle t^q\rangle$ as $q\rightarrow \theta$. This further emphasizes the transition in Eq.\,(\ref{eq:nuq}). The generic feature of Eq.\,(\ref{generalscal1}) is observed for variety of models as discussed in the companion paper \cite{Future_paper}, see also Fig.\,\ref{fig:sierpmom}.
\par The role of $\theta$ is most easily seen in the case of mean FPT $q=1$ of a particle diffusing in a wedge with an opening angle $\phi$. For $\phi<\pi/2$, the mean FPT for an infinite system is finite, and diverging as $\phi \rightarrow \pi/2$ \cite{redner2001guide}. As the persistence exponent is given by $\theta=\pi/2\phi$ \cite{redner2001guide}, Eq.\,(\ref{generalscal1}) indeed predicts that the mean FPT is independent of system size $L$ for $\phi<\pi/2$ but depends on $L$ for $\phi>\pi/2$. In the first case ($\phi<\pi/2$) the mean FPT is determined by the first scaling function $f_{\infty}(\cdot)$, whereas for $\phi>\pi/2$ it is determined by $\mathcal{I}(\cdot)$. This scenario highlights the need of the two different scaling regions. In general, Eq.\,(\ref{generalscal1}) predicts that when $\theta<1$ the mean FPT is determined by the rare fluctuations, while for $\theta>1$ it is described by typical fluctuations which are $L$ independent.

\par
\begin{figure}[t]
    \centering
    \includegraphics[width=0.5\textwidth]{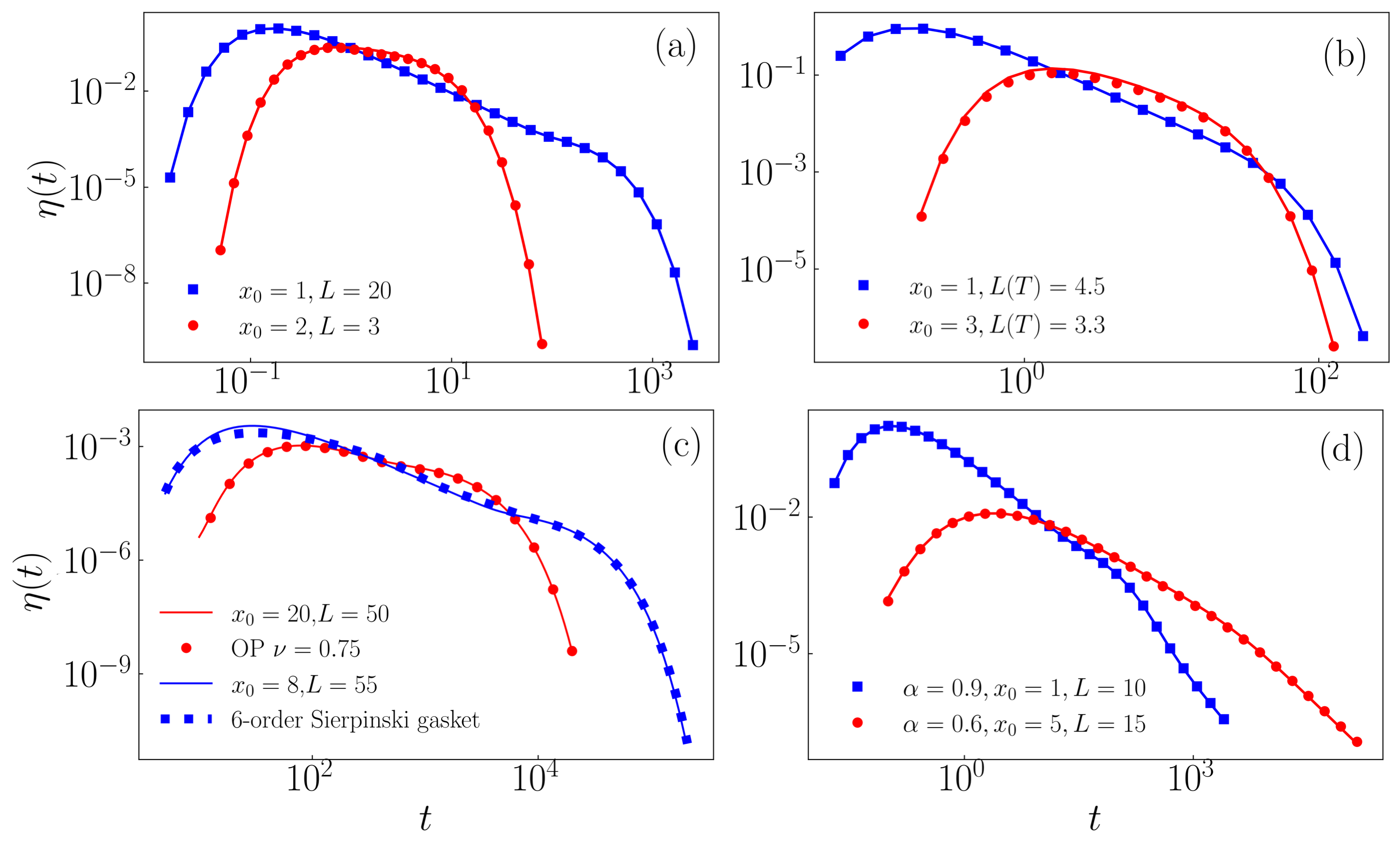}
    \caption{Theory of FPT PDF for compact domains (solid lines), Eq.\,(\ref{eq:uniform2}), perfectly matches numerical results  (symbols) for different models without fitting: (a) particle in a box, (b) Harmonic oscillator, (c) Sierpinski gasket, $d_f = \log(3)/\log(2)$ and $d_w = \log(5)/\log(2)$ (blue squares) and bi-scaling solution (smooth curve) found using the OP approach. Fractal geometry - OP model (red circles) with $d_f=1.5, d_w=2$ for finite system and the asymptotic bi-scaling theory. (d) CTRW. More details on the numerical data is given in SM.}
    \label{fig:unifrom}
\end{figure}

 \textit{Particle in a box}-- Consider the simple example of a diffusing particle in an infinite potential well taking place in the domain $(0,L]$, starting at position $x_0$. Here, $\tau_0=x_0^2/D$ and $\tau_L=L^2/D$, where $D$  is the diffusion constant, with our analysis focusing on the large-system limit, specifically $\tau_L \gg \tau_0$. The FPT PDF $\eta(t)$ is determined by the probability current at the absorbing boundary $x=0$ \cite{redner2001guide}, expressed as $\eta(t)=-D\partial_x P(x,t)|_{x=0}$, with $P(x,t)$ being the solution of the diffusion equation. The evaluation of $\mathcal{I}(\cdot)$ is achieved in the accompanying paper \cite{Future_paper} using a Mellin transform. From Eq.\,(\ref{definfhar}) with $\theta=1/2$ and $d_w=2$, we find
\begin{equation}\label{eq:inf1d} \mathcal{I}_{box}(\tau)= -\partial_{\tau }\vartheta_2\left(e^{-\pi^2 \tau}\right)\, ,
\end{equation}
 where $\vartheta_2(\cdot)$ is the Jacobi elliptic theta function of order $2$. An examination of $\mathcal{I}(\tau)$ reveals that, due to its power-law behavior in the small $\tau$ limit $\mathcal{I}(\tau) \propto \tau^{-3/2}$, this function is not normalized and, therefore, cannot be considered a PDF. Nevertheless, it still can be used to find all the moments of order $q>1/2$, as predicted in Eq.\,(\ref{generalscal1}). The short-time limit of $\mathcal{I}(\cdot)$ strikingly coincides with the long-time asymptotics of the FPT PDF for an infinite system, as given by the L\'evy-Smirnov-Schrödinger function  \cite{schrodinger1915theorie}
\begin{equation}\label{eq:free1d}
    f_{\infty}^{1d}(t) = \frac{x_0}{\sqrt{4\pi D}}t^{-3/2}\,e^{-x_0^2/4Dt}\underset{t \gg x_0^2/D}{\propto} t^{-3/2} \, ,
\end{equation}
leading to $\theta=1/2$. Consequently, when $L$ is large though finite, the PDF of the FPTs is predominantly described by the solution for the infinite system during short times. On the other hand, at longer times, the function $\mathcal{I(\tau)}$ in Eq.\,(\ref{eq:inf1d}) exhibits the effect of confinement due to its dependence on $\tau =t/\tau_L$. For intermediate times, $\tau_0 \ll t\ll \tau_L$, the expressions for those functions match, hence Eq.\,(\ref{eq:uniform2}) remains valid for large systems across both short and long time regimes, as illustrated in Fig. \ref{fig:unifrom}(a).
\par We will soon extend this result to models of diffusion on fractal geometries. However, we first focus on a widely applicable scenario in which a binding force field acts on the particle, coupled to a thermal bath. In this case, we identify the scale \(L\) with one derived from the Boltzmann steady-state measure, making it temperature-dependent. While the following example illustrates a thermal steady state, we demonstrate in the accompanying paper that active transport, which breaks detailed balance, and systems with resetting also exhibit bi-scaling behaviors \cite{Future_paper}. Importantly, the functions \(I(\cdot)\) and \(f_{\infty}(\cdot)\) are calculated explicitly. Thus, although identifying the diverging length scale \(L\) requires some basic physical insight, the theory applies to scenarios both with and without an equilibrium steady state.

\par \textit{Particle in a force field}-- We consider the dynamics of a one-dimensional Brownian particle influenced by a confining potential $V(x)$, with an absorbing boundary at $x=0$. The evolution of the propagator \(P(x,t)\), representing the probability of finding the particle at position \(x\) at time \(t\), is governed by the Fokker-Planck equation 
\begin{equation}\label{eq:fp1d}
\frac{\partial}{\partial t}P(x,t) = \tilde{\mathcal{L}}_{FP} P(x,t),
\end{equation}
where the Fokker-Planck operator \(\tilde{\mathcal{L}}_{FP}\) is defined as
\begin{equation}\label{eq:lfp}
\tilde{\mathcal{L}}_{FP} = D\left\{\frac{\partial^2 }{\partial x^2} + \frac{1}{T}\frac{\partial}{\partial x}\left(\frac{\partial V(x)}{\partial x}\right)\right\},
\end{equation}
and $T$ is the temperature of the reservoir and the Boltzmann constant is set to one.
\par Our first aim is to show that for a general form of $V(x)$, in the high-temperature limit the mean FPT is sensitive only to the large $x$ behavior of $V(x)$. In the case of a general potential of the form \( V(x) \sim  \kappa_{\beta} |x|^\beta + \kappa_{\beta-1} |x|^{\beta-1} + \ldots \) for large \(|x|\), with \(\beta > 1\), we find that \( \langle\, t\, \rangle \sim x_0  T^{1/\beta} \kappa_{\beta}^{-1/\beta} \Gamma(1 + \beta^{-1})/D + O(T^{-2/\beta}) \) \cite{Future_paper}. This is a special case of Eq.\,(\ref{generalscalfrac}) with $d_w=2,\theta=1/2$ where $q=1$. Here, as the confinement arises from the potential rather then the boundaries, we find the the length scale $L$ in Eq.\,(\ref{eq:nuq},\ref{generalscalfrac}) is given by \(L=(T/\kappa_{\beta})^{1/\beta}\sqrt{\Gamma(3/\beta)/\Gamma(1/\beta)}\) where $T$ is large. Two key examples include the V-shaped potential and harmonic potential, corresponding to $\beta =1$ and $\beta=2$, respectively. To describe the long-time limit we find from Eq.\,(\ref{definfhar}) \cite{Future_paper}
$ \mathcal{I}_{Harmonic}(\tau)= (1/\sqrt{8\pi} )e^{-\tau/4}\left(1-e^{-\tau/2}\right)^{-3/2}\, $, 
and $\mathcal{I}_{V-shaped}(\tau) = (1/\sqrt{4\pi})\tau^{-3/2}e^{-\tau}$ .
While combined with Eq.\,(\ref{eq:free1d}), those functions provides the full description of the FPT PDF for all $t$ as given in Eq.\,(\ref{eq:uniform2}). Remarkably, in both cases $\mathcal{I}(\tau)\propto \tau^{-3/2}$ and thus similar to the particle in a box, $\mathcal{I}(\cdot)$ is non normalizable and hence not a PDF. One can show that these solutions which describe long FPTs, match the L\'evy-Smirnov-Schrödinger function in Eq.\,(\ref{eq:free1d}), and thus we may construct a global solution to the problem using Eq.\,(\ref{eq:uniform2}), with $\theta=1/2$. Physically when $L$ is large compared to $x_0$, at short times the effect of the force is negligible, while at large FPTs, one needs to use $\mathcal{I}(\cdot)$.

\par In Fig.\,\ref{fig:sierpmom} we present the moments $\langle t^q \rangle$ versus $L$ for the FPT $t$ in the V-shaped potential. We see, in agreement with Eq.\,(\ref{eq1}), that the moments of the FPT are $L$ independent for $q$ values below the critical threshold, while this does not hold true for larger $q$. The transition's critical value, is the case when $q$ is equal to the persistence exponent $q=\theta=1/2$ as our theory predicts. The figure demonstrates how data analysis provides a clear clue for the failure of simple mono-scaling theories and a statistical transition taking place when $q=\theta$. 
\par For a general value of $\beta>1$, the solution to Eq.\,(\ref{eq:fp1d}) is derived using an eigenfunction expansion \(P(x,t) = e^{-V(x)/2+V(x_0)/2}\sum_n \psi(x_0) \psi_n(x) e^{-\tilde{\lambda}_n t/L^2}\), in which $e^{-V(x)/2}\psi_n(x)$ represent the eigenfunctions of the Fokker-Planck operator outlined in Eq.\,(\ref{eq:lfp}) \cite{risken1996fokker}. These eigenfunctions are normalized such that $\int_0^{\infty} |\psi_n(x)|^2dx = 1$. One can show that the eigenvalues \(\lambda\) scale as \(\lambda \propto L^{-2}\) for large $L$. Writing \(\lambda \equiv \hat{\lambda} L^{-2}\). By applying the limit introduced in Eq.\,(\ref{definfhar}), we can write the scaling function $\mathcal{I}(\cdot)$ as
\begin{equation}\label{eq:geniforce}
    \mathcal{I}(\tau)=\sum_n |\psi_n'(0)|^2 e^{-\hat{\lambda}_n \tau}\,,
\end{equation}
where $\tau=t/\tau_L$ and $\tau=L^2/D$ as mentioned. This function provides the long-time statistics of the FPT, namely $\tau\gg 1$, in the large-$L$ limit. Irrespective of the potential's particular characteristics, it is demonstrated that in the realm of short time intervals—characterized by predominately large eigenvalues —the resulting function aligns with the solution for a free particle given in Eq.\,(\ref{eq:free1d}), as detailed in the the companion paper \cite{Future_paper}. The significance of this finding is that for any binding field with \(\beta>1\), the bi-scaling theory works well, and the function \(\mathcal{I}(\tau)\) is independent of the initial condition, unlike \(f_{\infty}(t)\).

\par \textit{Fractal geometry}-- We next consider random walks on fractals, which are widely studied due to their relevance to understanding anomalous diffusion and diffusion in disordered media \cite{Havlin1987Diffusion,Reis2019Models}. Depending on the geometry of the fractal, different values of $d_w$ and $d_f$ emerge, for example, for the Sierpinski gasket $d_f = \log(3)/\log(2)$ and $d_w = \log(5)/\log(2)$ \cite{wu2011random}. Note that we limit our study to the case of compact processes, namely $d_w>d_f$ \cite{ben2000diffusion}. FPTs in such structures are often studied using the well-known mean field O'Shaughnessy-Procaccia (OP) transport equation \cite{o1985diffusion,o1985analytical,condamin2007first}. For an infinite system the function $f_{\infty}(\cdot)$ is known \cite{fa2003power,meroz2011distribution} and gives in the long time limit $f_{\infty}(t)\propto t^{-2+d_f/d_w}$, namely the persistence exponent is $\theta=1-d_f/d_w$. Meyer et al  \cite{meyer2011universality} found the integer moments of the process $|t|^q$ where $q=1,2,3,4,5$ and identified a scaling function that provides these moments. Note that in all these cases $q>\theta$. Interestingly, they noted this function is not normalized, and this observation was the starting point of our research. The emergence of a non-normalized solution that still yields the integer moments appears paradoxical at first, as moments are typically derived from normalizable distributions. We solved this open problem by employing the bi-scaling theory in Eq.\,(\ref{eq:uniform2}), namely both $f_{\infty}(\cdot)$ and $\mathcal{I}(\cdot)$ are needed to provide a global solution to the problem. The functions $f_{\infty}(\cdot)$ and $\mathcal{I}(\cdot)$ are given explicitly in the companion paper \cite{Future_paper}. Importantly, similar to other models, the divergence of $\mathcal{I}(\cdot)$ for small $\tau$ aligns with the large time behavior of $f_{\infty}(\cdot)$, allowing us to use Eq.\,(\ref{eq:uniform2}).

To validate these results, we conducted numerical simulations presented in Fig.\,\ref{fig:unifrom}(c) for diffusion on the Sierpinski gasket. We also compared our analytical findings with numerical calculations of the mean field OP model for different values of $d_f$ and $d_w$. As a result of the bi-scaling feature found here, the resulting PDFs can not be collapsed on a single master curve. This issue is discussed in further detail in our companion paper \cite{Future_paper}, but roughly speaking, the behavior at short times shown in Fig.\,\ref{fig:unifrom} is $L$ independent, while for long times, it depends on the system size, and hence the mono-scaling theory fails.

\par \textit{CTRW }--
The earlier findings presented in this work describe Markovian processes. However, diffusion in disordered systems, where a particle is hopping between traps, is often modeled by the CTRW \cite{metzler1994fractional,kenkre1973generalized,metzler2000random}. We consider a one-dimensional random walker, with waiting times between jumps are taken from the PDF $\varphi(t)\sim \alpha t^{-1-\alpha}/|\Gamma(-\alpha)|$, with $\alpha\in (0,1)$ where each jump is independent of the previous jumps. The CTRW is a non-Markovian model, and its limit is described by a fractional time Fokker-Planck equation \cite{metzler2000random,barkai2001fractional}
 \begin{equation}\label{eq:ffp}
    \frac{\partial }{\partial t}P(x,t)=K_{\alpha}\text{ }_{0}{\mathcal{D}_{t}^{1-\alpha}} \tilde{\mathcal{L}}_{FP} P(x,t),
\end{equation}
 where $K_{\alpha}$ is a generalized diffusion coefficient, $\tilde{\mathcal{L}}_{FP}$ is defined in Eq.\,(\ref{eq:lfp}) with $D=1$, and $\text{ }_{0}{\mathcal{D}_{t}^{1-\alpha}}$ is the fractional Riemann-Liouville operator \cite{metzler2000random}. 
The limit for the infinite system's PDF is given by $f_{\infty}(t)\propto t^{-1-\alpha/2}$ \cite{barkai2001fractional,balakrishnan1985anomalous}, namely $\theta=\alpha/2$, and further $d_w=2/\alpha$. Our findings reveal that despite the non-Markovian nature of the process, we observe a comparable bi-scaling pattern in $\eta(t)$ as seen in prior cases, with one significant distinction, as we shall discuss shortly. For CTRW in a force field, we solve Eq.\,(\ref{eq:ffp}) by employing the known subordination technique \cite{kolokoltsov2009generalized,barkai2000continuous,gorenflo2007continuous}, where roughly speaking exponentials are replaced with Mittag-Leffler functions \cite{erdelyi1953higher}. As in previous analyses, we consider the scenario where both $t$ and $L$ approach infinity, yet the ratio $t/L^{1/\alpha}$ remains constant.
\par For the example of a particle in a box, we find that $\mathcal{I}(\tau)$ exhibits two different power-law behaviors: one in the short-time limit, $ \mathcal{I}(\tau) \sim (\alpha/2)\, \tau^{-1-\alpha/2}/\Gamma(1-\alpha/2)$, 
 while another power-law emerges in the long-time limit $ \mathcal{I}(\tau) \sim \alpha\, \tau^{-1-\alpha}/\Gamma(1-\alpha)$. This is different then the previous cases where $\mathcal{I}(\tau)$ exhibits exponential decay in the long-time limit. Clearly this is an effect of the fat-tail trapping-time statistics. As seen in Fig. \ref{fig:unifrom}(d), Eq.\,(\ref{eq:uniform2}) provides the global solution for $\eta(t)$ for all $t$, with excellent agreement with the numerical calculation without fitting.

\par \textit{Discussion and final remarks}-- 
 We solved the problem of the scaling properties of first passage times for compact search by applying a bi-scaling theory which unified two scaling behaviors. We find it hinges upon two scaling functions, which we have derived for nine widely applicable models \cite{Future_paper}. The generality of our approach, including the transition of the moments at $q=\theta$ seen in Eq.\,(\ref{generalscalfrac}), and the fact that the scaling function $\mathcal{I}(\tau)$ does not depend on the initial condition, is demonstrated by the theory's applicability across diverse models and physical scenarios, as shown in Fig.\,\ref{fig:unifrom}. In some cases the length scale $L$, whether associated with system size, temperature, or typical force, does not need to be excessively large for the theory to remain effective.

\par Furthermore, the implications of our findings extend beyond single-agent dynamics. One method to speed up hitting times is by using many non-interacting agents
 searching for the target, for example this strategy is used in biology
 when sperms search for an egg \cite{sposini2024being,grebenkov2022reversible,linn2022extreme,lawley2023slowest}.
 In a nutshell, one searcher out of many, either by an inherent advantage or by luck, will be by far faster locating the target than typical searchers, hence hordes can speed up the search time tremendously. Our work sheds new light on this important problem. In these scenarios, the statistics of the fastest particles are governed by \(f_{\infty}(\cdot)\), whereas the slowest particles are characterized by \(\mathcal{I}(\cdot)\), whose role gain importance when the number of particles in the system is increased.

\par Finally, the variations in the shapes of $\eta(t)$ from one model to another potentially explain the absence, prior to our work, of a general theory for this observed phenomenon. For example, for long times, the decay of $\eta(t)$ can exhibit either exponential or power-law behavior, where the second corresponds to the case of strong disorder as modeled by CTRW with $\alpha <1$. Notwithstanding these rich types of behaviors, our theory provides a unified framework to capture all of them.

\par \textbf{Acknowledgments}-- The support of Israel Science Foundation's grant 1614/21 is acknowledged. We would like to thank Olivier B\'enichou  and L\'eo R\'egnier for useful discussions.

\end{document}


\title{SUPPLEMENTARY MATERIAL}

\maketitle

\section{Figure preparation}
To prepare Fig. 3 in the Letter, we employed the analytical expressions provided in Table 2 with Eq. (4) in the main text. The exact solutions for Figs. 3(a,b,d-f) were obtained using the eigenfunction expansion method, as elaborated in the companion paper. Fig. 3(c) was generated using the solution presented in the third line of Table 2. To ensure convergence to the solution, we summed the eigenfunctions up to order $n\geq 100$. Specifically, for Figs. 3(a-d), we set $D=1$, while for Fig. 3(f), we used $K_{\alpha}=1$ with $\alpha=0.6,0.9$. 
 Here is a breakdown of the figure-specific details:
\begin{itemize}
\item  Fig. 3(a): The theoretical prediction, as outlined in the first line of Table 2, was computed using the Jacobi elliptic theta function of order 2, represented as $\vartheta_2(.)$, and implemented through the Python library Mpmath.

\item  Fig. 3(d): To generate the plot associated with the expression found in the fourth line of Table 2, we computed the values of both $k$ and $c_k$ as defined in the fourth line of table 2, directly using Python libraries, including NumPy and SciPy.

\item Fig. 3(e): For the numerical computation of the O'Shaughnessy-Procaccia (OP) model (represented by red symbols), we considered the fractal dimension and the walk dimension to be $d_f=1.5$ and $d_w=2$, respectively. As for the Sierpinski gasket, which has a fractal dimension of $d_f=\log(3)/\log(2)$ and a walk dimension of $d_w=\log(5)/\log(2)$, the outcomes pertaining to the FPT distribution were obtained through numerical calculations, as elaborated below in the following section.

The characteristic length $r_0$ is determined by $r_0 = n^{1/d_f}$ where $n$ is the number of nodes within radius $r$, which signifies the chemical distance between the initial position and the target. The system's size $L$ is defined similarly, with $n$ indicating the number of nodes within the entire graph.

\item  Fig. 3(f): Generating the theoretical predictions given in the sixth line of Table 2, we made use of the general Mittag-Leffler function denoted as  \cite{erdelyi1953higher}
\begin{equation}\label{MLdef}
E_{\alpha,\alpha}(z) =\sum_{k=0}^\infty \frac{z^k}{\Gamma(\alpha k+\alpha)} . 
\end{equation}
Additionally, we employed the one-sided L\'evy functions, denoted as $l_\alpha(t)$. These L\'evy functions are defined in Laplace space through the integral \begin{eqnarray}\label{eq:laplevy}
    \hat{ \ell_{\alpha}}(u) =\int_0^{\infty}\exp(-ut)l_\alpha(t)dt =  e^{-u^{\alpha}}.
\end{eqnarray}
Both of these functions are readily accessible through computational tools such as Mathematica.

\end{itemize}
\par To prepare Fig.\,2 in the main text, we used the FPT PDF for the V-shaped potential, specifically $F(x) = \text{sgn}(x)F_0 $.
\begin{enumerate}
    \item The symbols are given by the exact moments; For a particle that initiates its motion at $x=x_0$ with an absorbing boundary at $x=0$, the FPT PDF is given by \cite{schrodinger1915theorie}:
\begin{equation} 
    \eta(t)=\frac{x_0}{2\sqrt{D\pi}}t^{-3/2}e^{\frac{-(x_0-D\frac{F_0}{k_B T}t)^2}{4Dt}} \, .
\end{equation}
From which the FPT moments are calculated by $\langle t^q \rangle = \int_0^\infty t^q\eta(t)dt$, gives
 \begin{equation}
    \langle t^q\rangle=\frac{\tau_0^{\frac{q}{2}+\frac{1}{4}}}{\sqrt{\pi}} \tau_L^{\frac{q}{2}-\frac{1}{4}} e^{\sqrt{\frac{\tau_0}{4\tau_L}}} K_{\frac{1}{2}-q}(\sqrt{\frac{\tau_0}{4\tau_L}})\, ,
\end{equation}
where $K_n(.)$ is the modified Bessel function of the second kind. 
\item The dashed lines are the theory; Using the scaling functions given in the third line of table 2, we obtain the moments for $q<1/2$ using 
\begin{equation}
    \langle t^q\rangle_{q<1/2} = \int_0^\infty t^qf_{\infty}(t)dt=\frac{4^{-q}}{\sqrt{\pi}}\Gamma(\frac{1}{2}-q)x_0^{2q}
\end{equation}
and for $q>1/2$ using 
\begin{equation}
    \langle t^q\rangle_{q>1/2} = L^{2q-1}x_0\int_0^\infty \tau^q\mathcal{I}(\tau)d\tau=\frac{4^{q-1}}{\sqrt{\pi}}\Gamma(q-\frac{1}{2})x_0 L^{2q-1}
\end{equation}
with $x_0=1$.

\end{enumerate}

\begin{table}[h]
\centering
\begin{tabular}{||c c c c c c||} 
 \hline
 model & $d_f$ & $d_w$ & $\theta$ &$\tau_0$& $\tau_L$ \\ [0.5ex] 
 \hline\hline
 1d box & 1 & 2 & 1/2 &$x_0^2/D$ &$L^2/D$\\ 
 harmonic potential & 1 & 2 & 1/2 &$x_0^2/D$ &$ k_B T/4\kappa D$ \\
 V-shaped potential & 1 & 2 & 1/2 & $x_0^2/D$& $4 k_B^2T^2/F^2 D$\\
  wedge & 2 & 2 & $\nu/2$  &$r_0^2/D$ &$L^2/D$\\
   OP fractal & $d_f$ & $d_w$ & $1-d_f/d_w $  &$r_0^{d_w}/K$ & $L^{d_w}/K$\\

 CTRW & 1 & $2/\alpha$ & $\alpha/2$  &$(x_0^2/K_{\alpha})^{1/\alpha}$ & $(L^2/K_{\alpha})^{1/\alpha}$\\
 [1ex] 
 \hline
\end{tabular}
\caption{Systems exponents and scaling parameters for the different models.}
\label{table:1}
\end{table}

\begin{table}[t]
\centering
\renewcommand{\arraystretch}{1.5}
\begin{tabular}{@{}ll@{}ll@{}}
\toprule
\textbf{Model} & \textbf{$f_{\infty}(t)$}& \textbf{$\mathcal{I}(\tau)$} \\ 
\midrule
1d box & $(x_0/\sqrt{4\pi D}) t^{-3/2}\exp\left(-\frac{x_0^2}{4D t}\right)$ &  $-\partial_{t }\vartheta_2\left(e^{-\pi^2  \tau}\right)$ \\
Harmonic force & $(x_0/\sqrt{4\pi D})t^{-3/2}\exp\left(-\frac{x_0^2}{4D t}\right)$ & $\sqrt{\frac{1}{8\pi}} \frac{ e^{-\tau/4}}{(1-e^{-\tau/2})^{3/2}}$ \\ 
V-potential & $(x_0/\sqrt{4\pi D}) t^{-3/2}\exp\left(-\frac{x_0^2}{4D t}\right)$ & $\sqrt{\frac{1}{4\pi}} \tau^{-3/2}e^{-\tau} $\\
\addlinespace
Wedge & $ \frac{\nu r_0}{4 t^{3/2}}e^{-r_0^2/4Dt}$ &$\frac{2^{-\nu}}{\Gamma(1+\nu)}\sum_k c_k k^{2+\nu}  e^{-k^2 \tau} $\\
& $\left( I_{\frac{3+\nu}{2}}(\frac{r_0^2}{8 D t}) - (1-\frac{8Dt}{r_0^2}(1+\nu))I_{\frac{1+\nu}{2}}(\frac{r_0^2}{8 D t}) \right) $ \\
\addlinespace
Fractal & $ \frac{d_w^{-2+2\nu} }{2 \Gamma(\nu) } e^{-\frac{\tau_0}{d_w^2t}}( t/\tau_0)^{-2+\nu}  /\tau_0$
& $ \frac{d_w 2^{2\nu-3} \Gamma(\nu) }{\Gamma(2-\nu)} \sum_{n=0}^{\infty}  \frac{J_{\nu}(z_{-\nu,n}) }{J_{1-\nu}(z_{-\nu,n})} z_{-\nu,n}^{3-2\nu}  e^{-d_w^2 z_{-\nu,n}^2 \tau/4}$\\
\addlinespace
CTRW & $(\frac{K_{\alpha}}{x_0^2})^{1/\alpha}\ell_{\alpha/2}\left(\frac{t}{x_0^{2/\alpha}}\right) $
& $  \sum_{n=0}^\infty 2\pi^2\left(n+\frac{1}{2}\right)^2 \tau^{\alpha-1}E_{\alpha,\alpha}\left(-\pi^2\left(n+\frac{1}{2}\right)^2 \tau^{\alpha}\right)$\\
\bottomrule
\end{tabular}
\caption{The scaling functions $f_{\infty}(t)$ and $\mathcal{I}(\tau)$ for each of the models discussed in this work. For the 2D wedge - $I_{\nu}(.)$ denotes the modified Bessel of the first kind, $k$ is the solution of $J_{\nu+1}(k)=J_{\nu-1}(k)$ and $c_k$ is given by $c_k=\int_0^1 r J_{\nu}(k r) dr/\int_0^1 r J_{\nu}^2(k r) dr$. For the fractal - $z_{-\nu,n}$ is the $n$-th zero of the $-\nu$ order Bessel function, K is the diffusion coefficient for the OP equation, and $\mu=d_f/d_w$. For CTRW -  $E_{\alpha,\alpha}(.)$ is the Mittag-Leffler function, defined as $E_{\alpha,\alpha}(z)=\sum_{k=0}^{\infty}z^k/\Gamma(\alpha k+\alpha)$. }
\label{table:2}
\end{table}
\section{Simulation on diffusion on the Sierpinski Gasket}\label{SM:simulation}
 
We employ an exact enumeration technique, which we will now elaborate on. The master equation, which governs the probability $P_{n,i}$ for a random walker on a graph denoted as $G$ to be at site $i$ at step $n$, is generally given as follows:

\begin{equation}
    P_{n+1,i} = \sum_j M_{i,j} {P}_{n,j},
\end{equation}
where $M$ is the transition matrix, namely the transition probability from site $i$ to site $j$ is given by the matrix element $M_{j,i}$. To model an absorbing boundary condition at the target site $i=i_{\text{target}}$, we impose $P_{n,i_{\text{target}}} = 0$. The survival probability, which quantifies the likelihood that the particle has not reached the target by the $n$-th step, is defined as:

\begin{equation}\label{eq:surv_sierp}
    S_n = \sum_{i} P_{n,i}.
\end{equation}
The FPT probability distribution $\eta_n$ is then determined by computing the numerical derivative of Eq. (\ref{eq:surv_sierp}), which is expressed as:
\begin{equation}
    \eta_n = \frac{S_{n} - S_{n-1}}{\Delta n}, 
\end{equation}
where $\Delta n=1$. In the context of fractal geometry, we set $L = N^{1/d_f}$, where $N$ represents the number of nodes in the graph. The relationship between the number of nodes $N$ and the fractal order $m$ (indicating the number of iterations or recursive steps taken to generate the fractal pattern) for the Sierpinski gasket can be described generally as $N=(3 + 3^{m+1})/2$ \cite{wu2011random}. To ensure consistency with the large-system limit, we deliberately position the source and target points in close proximity to each other to ensure $r_0 \ll L$. However, as depicted in Fig. 3(e) in the main text, the theory remains effective even when \(L\) is not necessarily significantly larger than the initial distance from the target, similar to the behavior observed in other models.